\documentclass[aps,prb,twocolumn,superscriptaddress,showpacs]{revtex4-1}

\usepackage{amsmath}
\usepackage{graphicx}
\usepackage{calc}
\usepackage{bm}

\usepackage[T2A]{fontenc}
\usepackage[cp1251]{inputenc}
\usepackage[english]{babel}

\begin{document}

\title{Memory effect in the charge transport in strongly disordered antimony films}

\author{N.N. Orlova}
\email[E-mail:~]{honna@issp.ac.ru}

\author{S.I. Bozhko}
\author{E.V. Deviatov}

\affiliation{Institute of Solide State Physics Russian Academy of Sciences, 142432, Chernogolovka, Moscow District, Russia}

\date{\today}

\begin{abstract}
We study  conductivity of strongly disordered amorphous antimony films  under high bias voltages.  We observe non-linear current-voltage characteristic, where the conductivity value at  zero bias is one of two distinct values, being determined by the sign of previously applied voltage. Relaxation curves demonstrate high stability of these  conductivity values on a large timescale. Investigations of the antimony film structure allows to determine the percolation character of electron transport in  strongly disordered films. We connect the memory effect in conductivity with modification of the percolation pattern due to recharging of some film regions at high bias voltages. 
\end{abstract}


\maketitle

\section{Introduction}

Recently, considerable interest is attracted by structures with a memory effect, where  the electrical resistance  depends on the measurement backstory~\cite{mem1,mem2}. In many cases, e.g. for  MOSFETs, the resistance of the dielectric layer can be irreversibly affected by high gate voltages, so this new dielectric state maintains after  removing the gate voltage. On the other hand, structures are possible in which the resistance  at zero bias is controlled by the sign/value of the previously transmitted current. For the first time such structures were considered theoretically~\cite{chua} as the fourth missing element of electrical circuits, and were called as a "memristor". Although there are some doubts about practical implementation of the memristor model~\cite{ventra}, this topic is attracting significant interest not only due to the practical  potential for signal processing,  storing information~\cite{inf}, but also due to  the general interest to  physical processes which lead to reversible resistance change in various structures.

There are two main approaches to the practical realization of  resistance memory effects: (i) three-layer vertical metal-dielectric-metal structures, where the resistance is determined by ion diffusion within the dielectric layer (see, e.g., Refs~\onlinecite{tulina,vedeneev}  and references herein); (ii) systems with  phase transition, which occurs between crystalline and amorphous states under the control current pulse (see, e.g., Ref.~\onlinecite{phase} as a review).  Well known drawbacks of these systems are the relatively slow performance rate, and also the low lifetime (about 100 cycles) in the first case because of the poorly  controlled ion diffusion process.    

On the over hand,  there are many important results in the field of  electron transport of strongly disordered systems near the metal-dielectric transition~\cite{dolgop,gantm}. The effects observed, e.g., evolution  of the current percolation pattern~\cite{shklovskii} under the influence of external factors, can be the basis for novel realizations of  resistive memory effects. In this case, the restrictions are reduced on the low switching rate and  sample survival in successive cycles due to electron transport instead of the ion one. This topic is also of general physical interest, since observation and investigation the resistive memory effects can provide addition information on charge transport in strongly disordered systems.  

Here,  we study  conductivity of strongly disordered amorphous antimony films  under high bias voltages.  We observe non-linear current-voltage characteristic, where the conductivity value at  zero bias is one of two distinct values, being determined by the sign of previously applied voltage. Relaxation curves demonstrate high stability of these  conductivity values on a large timescale. Investigations of the antimony film structure allows to determine the percolation character of electron transport in  strongly disordered films. We connect the memory effect in conductivity with modification of the percolation pattern due to recharging of some film regions at high bias voltages.

\section{Samples and techniques}

\begin{figure}[t]
\center{\includegraphics[width=0.8\columnwidth]{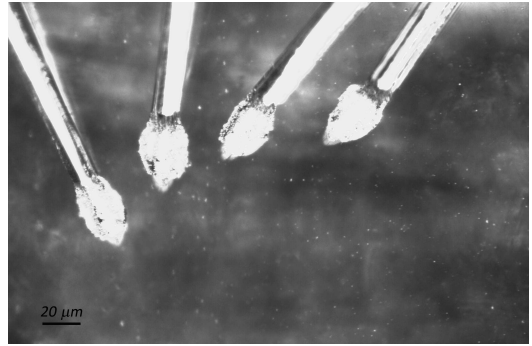}}
\caption{Optical image of the antimony film region with aluminum contacts. Measurements are performed in a two-point technique, voltage  $V$ is directly applied between two neighbor contacts. To obtain the differential conductivity $dI/dV(V)$, the applied voltage is modulated by a small ac component, a lock-in measures the ac component of the current $I$, which is proportional to the differential conductivity $dI/dV$.}
\label{sample}
\end{figure}

As for electron structure, antimony is a standard semimetal  with relatively low carrier concentration. It has several allotropic modifications in crystal and amorphous states, the conductivity of  amorphous and crystal antimony differs by four orders of magnitude~\cite{maki}. One of the amorphous modification is the black antimony, which can be produced by thermal evaporation on the amorphous substrate. The obtained film is amorphous for thicknesses below 10 nm, it has two-phase amorphous-crystalline structure for film thicknesses up to 120 nm, and it is a crystal for higher thicknesses. The exact critical thickness values depend on the deposition speed and the substrate temperature~\cite{hashimoto}. 

Our samples are the thin antimony films with 20~nm thickness. The film thickness and the evaporation parameters are chosen to have   strongly disordered state of the film in terms of both short- and long-range potential fluctuations. Antimony is deposited trough the cracking zone (820$^\circ$~C) on a glass (amorphous) substrate. The evaporation rate is 0.5~\AA/sec at the crucible temperature 440$^\circ$~C.  After deposition, the amorphous film structure is confirmed by x-ray diffraction method. As a reference sample, we use similarly evaporated film on a standard silicon wafer.

For transport measurements, we fabricate Ohmic contacts by conventional ultrasonic bonding of aluminum wires. This results in aluminium contacts of about  20~$\mu$m size. The bonding provides reliable destruction of any oxide/contamination layers between the   film and the wire materials, which results in good Ohmic contact. For amorphous films, cold bonding process has advantages over the standard lithographic methods, since there is no risk of  sample crystallization~\cite{fisher} due to essential sample  heating at the resist baking (90-130$^\circ$~C) and  contacts evaporation steps.

The contacts are placed in the central part of a lengthy film (approximately $0.5\times 0.5$~mm$^2$) at the distances of 20-30~$\mu$m between the contacts, see Fig.~\ref{sample}. This geometry allows to avoid any edge effects in distributions of electric fields and  current flow paths. In addition, it allows to study the structure of the film between the contacts by atomic force and scanning electron microscopy techniques. 

To apply high bias voltages to the sample (up to 10~V), the measurements are performed in a two-point technique: the bias voltage  $V$ is applied between the contacts, while the current  $I$ is measured in the circuit. One of the contacts in Fig.~\ref{sample} is grounded, the applied voltage varies within $\pm$10~V range at the neighbor contact, which corresponds to the maximum current of about 0.1~mA through the sample. To obtain  differential conductivity curves $dI/dV(V)$, the applied voltage is additionally modulated by a small (4.4~mV) ac component at a frequency of 1100 Hz. The ac current component is measured by lock-in, being proportional to differential conductivity $dI/dV$ at a given bias voltage $V$. We verify that the obtained $dI/dV$ value does not depend on  the modulation frequency in the range 100~Hz--1~kHz, which is determined by the applied filters. Due to the high quality of Ohmic contacts, their  resistances are much below the resistance of a disordered antimony film (more than 100~kOhm). Also, the contact quality is confirmed by measurements of a reference sample (which is the antimony film on a silicon surface), which demonstrates the resistance values less than 100~Ohm in a two-point technique. All the measurements are carried out at room temperature in zero magnetic field.

\section{Experimental results}

\begin{figure}[t]
\center{\includegraphics[width=\columnwidth]{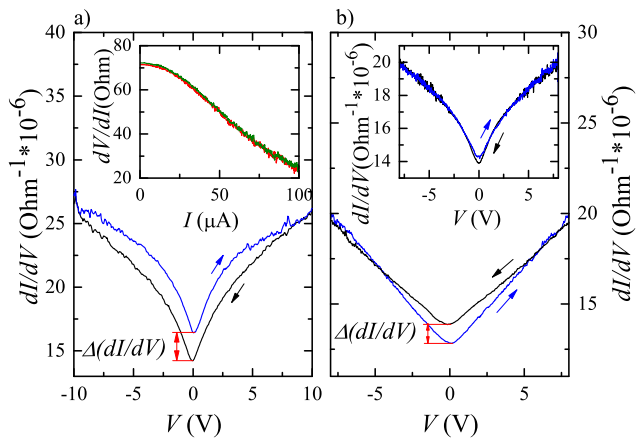}}
\caption{Dependence of differential conductivity $dI/dV(V)$  of a strongly disordered antimony film on the bias voltage for two different  samples, (a) and (b), respectively. The curves demonstrate strongly non-Ohmic behavior of $dI/dV(V)$, with hysteresis for two voltage sweep directions, an amplitude of the effect is $\Delta(dI/dV(V=0))=2.2\cdot 10^{-6}$~Ohm$^{-1}$ for (a). The inset to (a) shows the reference curve for the antimony film on a silicon substrate. The resistance difference in comparison with the main (a) panel well corresponds to the expected value~\cite{maki} (see the main text for details). The inset to (b) shows decrease of $\Delta(dI/dV(V=0))$ from $1\cdot 10^{-6}$~Ohm$^{-1}$ to $0.4\cdot 10^{-6}$~Ohm$^{-1}$ while reducing the sweep rate by an order of magnitude. Also, the curve shape is changed: the $dI/dV(V)$ curves coincide at high voltages $V$, while the nonzero difference $\Delta(dI/dV(V=0))$ can be clearly seen even for the lowest sweep rate.}
\label{IV}
\end{figure}

Fig.~\ref{IV} shows examples of experimental $dI/dV(V)$ curves, which demonstrate dependences of differential conductivity $dI/dV$ on the applied bias voltage $V$ for two different samples, in  (a) and (b),  respectively. In both the cases,  the 20~$\mu$m-long regions of antimony films  show large resistance at zero bias voltage (more than 100 kOhm) and strongly non-Ohmic behavior of $dI/dV(V)$. Namely, differential conductivity increases in about 2 times in a sweep of $V$ from zero to $\pm$10~V. Every curve shows a kink in the middle of the range, at higher biases  the differential conductivity $dI/dV$ is a linear function of $V$. The nonlinear behavior of differential  resistance is also observed for the reference sample of an antimony film on a polished oxidized silicon surface, see the inset to Fig.~\ref{IV} (a). However,  the measured value of differential resistance is significantly smaller (over three orders of magnitude) for the reference sample  for similar sample geometry. This resistance difference of two antimony films corresponds to the expected values for amorphous (on the glass substrate) and crystalline (on the silicon one) films~\cite{maki}: from the sample dimensions, the antimony resistivity can be estimated as 0.5~Ohm$\cdot$cm and $5\cdot10^{-4}$~Ohm$\cdot$cm, respectively.

To our surprise, there is a hysteresis in the experimental  $dI/dV(V)$ curves for two different sweep directions, see Fig.~\ref{IV}. Namely, if the linear parts of $dI/dV(V)$ curves are within the sweep range, the sample differential  conductivity at zero bias depends on the sweep direction. Qualitatively, this effect is well reproducible for different samples, cp. panels (a) and (b) in Fig.~\ref{IV}, while both the amplitude  $\Delta(dI/dV(V=0))$ and the sign of the effect vary from sample to sample. For a given sample, $dI/dV(V)$  curves are well reproducible in at least 100 voltage scans.  In addition, $\Delta(dI/dV(V=0))$ depends on the sweep rate: when it is decreased by an order of magnitude, the $dI/dV(V)$ curves coincide at high voltages $V$, while the nonzero difference $\Delta(dI/dV(V=0))$ can still be clearly seen even for the lowest sweep rate, see the inset to Fig.~\ref{IV} (b). We don't observe any hysteresis effects for the reference films on a silicon substrate. 

\begin{figure}[t]
\center{\includegraphics[width=\columnwidth]{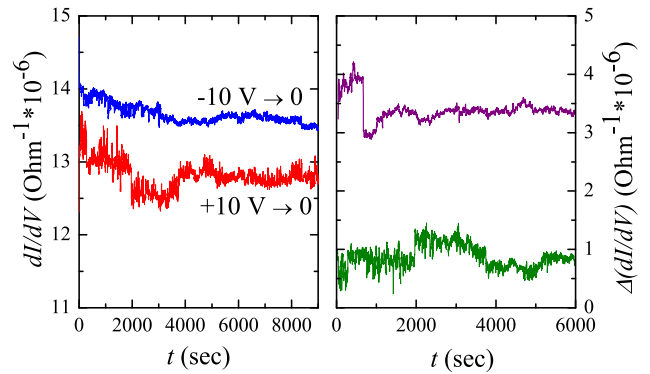}}
\caption{Time-dependent relaxation of differential conductivity at zero bias $dI/dV(V=0)$.  (a) Relaxation curves for the sample from Fig.~\ref{IV} (b),   after exposure at -10~V (blue curve) and at +10~V (red curve). (b) Time dependence for the hysteresis amplitude $\Delta(dI/dV(V=0))$ for two samples from Fig.~\ref{IV} (a) and (b), black and green curves, respectively. $\Delta(dI/dV(V=0))$  is obtained as a difference between the corresponding relaxation curves.}
\label{relax}
\end{figure}

The observed dependence of a hysteresis amplitude $\Delta(dI/dV(V=0))$ on the $V$ sweep rate  means that there are some relaxation processes in electron transport in strongly disordered antimony films. For this reason, we directly investigate relaxation  for our samples, see Fig.~\ref{relax} (a). To realize a procedure, we keep the sample for about 10 minutes  at high bias $\pm$10~V, which is obviously higher than the kink voltage value of $dI/dV(V)$ curves. After that we reset the bias to zero and immediately start tracing the difference conductivity $dI/dV(V=0)$ in dependence of time. The obtained relaxation curves show small (below 10\%) relaxation for the first 3000 seconds, the $dI/dV(V=0)$  value is stable afterward, see Fig.~\ref{relax} (a). This stable $dI/dV(V=0)$ value obviously depends on a sign of initial bias voltage. It well corresponds to the sign and the amplitude of the $dI/dV(V)$ hysteresis  for this sample in Fig.~\ref{IV} (b). Fig.~\ref{relax} (b) directly demonstrates  the hysteresis amplitude $\Delta(dI/dV(V=0))$ for two samples from Fig.~\ref{IV} (a) and (b), where $\Delta(dI/dV(V=0))$ is obtained by subtracting two relaxation curves for each sample.

\section{Discussion}

As a result, we demonstrate a memory effect for differential conductivity of a sample at zero bias voltage $dI/dV(V=0)$: there are two  distinct conductance values, switching between them is obtained by applying significant bias of a different sign. In addition, we show that these sample states  remain stable on  macroscopic time scales (several hours). It indicates that significant bias voltage applied between 20~$\mu$m  spaced contacts leads to a stable change of the current distribution pattern in strongly disordered antimony films. This process is well reproducible for different samples and in multiple cycles, i.e. sample switching is well reversible procedure.   

To explain the observed memory effect it should be noted, that in case of thin amorphous antimony film there are no reversible phase transitions between amorphous and crystalline states~\cite{fisher}. In other words, an amorphous antimony film can be transformed into a crystalline state of $\alpha$-antimony by heating, but the backward transition from a crystalline state into an amorphous one can not be made by melt quenching. Moreover, we observe a well developed memory effect at low sweep rates.  Therefore, phase transitions can not be considered in antimony films as a mechanism of the observed memory effect~\cite{phase}. Also, for the lateral sample geometry, the diffusion of dopant ions from electrodes can not be considered  for 20 $\mu$m-long  monocomponent  film, in contrast to thin  three-layer planar structures metal-insulator-metal~\cite{tulina}. In the latter case,  for a comparable potential difference at the electrodes $\pm$10~V, it drops on a 100~nm thick dielectric layer, which corresponds~\cite{vedeneev} to the $10^{6}$~V/cm electric field. In our case, the ion diffusion might only be possible from aluminum contacts  in their vicinity. Due to the high resistance of the film and low contact resistance (see the main figure and the insert to Fig.~\ref{IV} (a)), the contact areas  are of negligible contribution to the resistance of 20~$\mu$m-long sample. Moreover the average electric field $10^{4}$~V/cm is  too small for the diffusion process.

\begin{figure}[t]
\center{\includegraphics[width=\columnwidth]{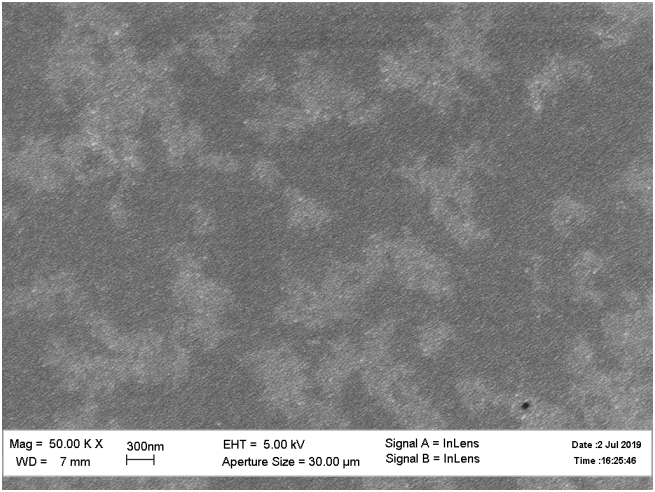}}
\caption{Electron microscope image of the investigated antimony film on a glass substrate. Coagulated conglomerates of deposited particles can be seen as light gray contrast areas with lengths from 300~nm to 1 $\mu$m, which well corresponds to a typical pattern of percolation clusters for strongly disordered materials~\cite{shklovskii}. The image is obtained with scanning electron microscope Zeiss Supra 50VP.}
\label{sem}
\end{figure}

On the over hand, the zero-bias sample resistance corresponds to resistivity $\approx$0.5~Ohm$\cdot$cm, so the film is in a dielectric state according to Ioffe-Regel criterion~\cite{gantm}.  Such strongly disordered thin films usually demonstrates percolation current patterns~\cite{shklovskii}, which could be affected by  significant bias voltage. Fig.~\ref{sem} shows the electron microscope image of the investigated antimony film on a glass substrate.  This film is of granular structure and shows coagulation of deposited particles, which is a fingerprint of amorphous antimony  films  obtained by low-rate deposition~\cite{palatnik}. Coagulated conglomerates of deposited particles can be seen as light gray contrast areas with lengths from 300~nm to 1 $\mu$m. Fig.~\ref{sem} well corresponds to a typical pattern of percolation clusters for strongly disordered materials~\cite{shklovskii}. For a reference film on a polished silicon wafer, crystalline phase dominates, which leads to low resistivity values without noticeable memory effects~\cite{butenko}.  

\begin{figure}[t]
\center{\includegraphics[width=\columnwidth]{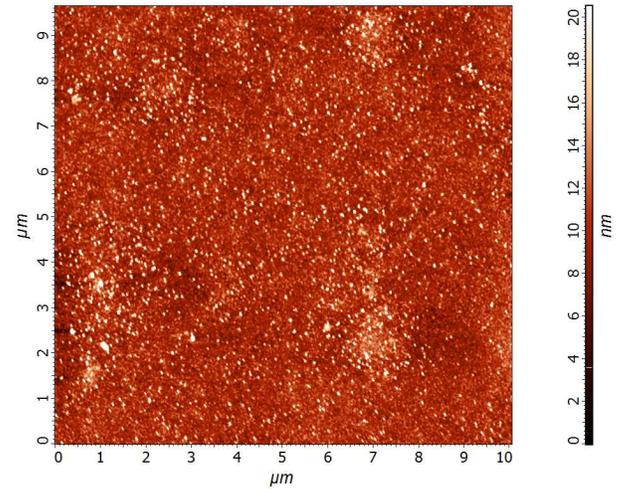}}
\caption{Atomic force image of the surface  of the investigated antimony film. The depicted in  Fig.~\ref{sem} film inhomogeneity  is connected with spatial variation of the film thickness, which leads to the  percolation structure for the current flow.}
\label{afm1}
\end{figure}

The film inhomogeneity in Fig.~\ref{sem} occurs due to the thickness variation around the film, which is demonstrated in Fig.~\ref{afm1} by atomic-force technique. The bottom layers of the film are filled completely, while the upper layer consists from the coagulated regions. The resulting thickness variation leads to the pattern of percolation clusters for the current flow. In this case the resistance between two sample contacts  is determined by several percolation channels, which are  connected in parallel. Thick sections of the sample in Fig.~\ref{sem} form the conducting clusters while the resistance is mostly determined by the critical regions of small thicknesses between these clusters~\cite{shklovskii}.

In the case of percolation current flow, conductivity is increasing at high  voltage biases, which is usually  explained by the breakdown of the resistive sections~\cite{shklovskii}. The conductivity  no longer depends on specific geometric dimensions of highly resistive area of a particular percolation channel when the field strength value $E=V/d$  (where $d\sim20$~$\mu$m is a sample length) allows for an electron to achieve the mobility edge by obtaining an energy $eEL_c$ at the characteristic localization length $L_c$. Because of the large number of clusters of different sizes in strongly disordered antimony films, there is no certain breakdown voltage for the experimental $dI/dV(V)$ curves, so we observe smooth increase of conductivity with voltage bias. At high biases, the conductivity becomes independent of the particular configuration of resistive regions, which leads to the coincidence of $dI/dV(V)$ curves at large $V$ values, as it can be seen in the inset to Fig.~\ref{IV} (b). 

\begin{figure}[t]
\center{\includegraphics[width=\columnwidth]{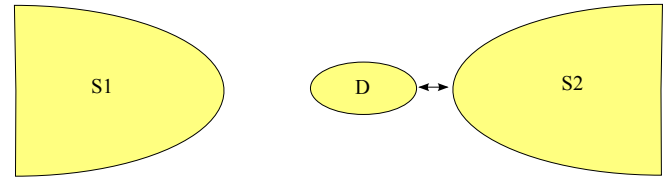}}
\caption{Schematic diagram of a small conductive cluster (D), separated by resistive junctions from the conductive regions (S1 and S2). The recharge process (marked by the arrow) occurs between the cluster and the nearest region S2.}
\label{discussion}
\end{figure}

The  described mechanism  is independent of a bias sign, which well corresponds to symmetric (i.e. even) $dI/dV(V)$ curves. On the other hand,  the existence of two distinct values of conductivity at zero bias  $dI/dV(V=0)$ indicates modification  of the percolation pattern. Since the modification is sensitive to the bias sign, it is only possible due to charging of some film regions~\cite{theory}.

Really, resistance of the percolation channel is determined by the most resistive area situated between two conductive regions~\cite{percolation}. For strongly disordered samples, for some of the channels, it is natural to expect that there is a small cluster (i.e. large quantum dot, electron trap) weakly coupled to the conductive regions, see Fig.~\ref{discussion}. The coupling strengths are obviously not equivalent  due to the arbitrary location of the cluster. When a high bias is applied between the regions S1 end S2, this cluster (trap) is charged or discharged, depending on the bias sign. The charging/discharging process goes through the breakdown to the nearest conductive region (D-S2 in Fig.~\ref{discussion}), so it requires finite breakdown voltage (see above). As a result, the cluster (trap) charging (or discharging) is accompanied by  redistribution of the electron density in the conductive regions, due to the screening effects~\cite{shklovskii}. Thus, the cluster (trap) state "charged/uncharged"  determines the configuration the most resistive area in the percolation channel, i.e. the value of the sample differential resistance at zero bias. Since the trap charging/discharging requires finite voltage, an actual configuration of the percolation channels is stable at zero bias, which is revealed in the relaxation curves in Fig.~\ref{relax}. In principle, similar recharge process has been predicted theoretically~\cite{theory}.

\begin{figure}[t]
\center{\includegraphics[width=\columnwidth]{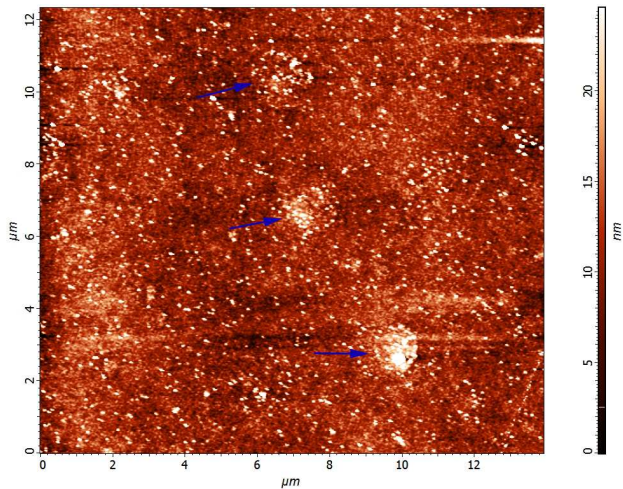}}
\caption{Atomic force image of  the film with inclusions of the crystalline phase (marked by blue arrows), which are located  one after another in the direction away from the contact (bottom of the image). These inclusions can play the role of rechargeable traps, which determine the configuration of resistive regions in the sample, i.e. the sample resistance value at zero bias voltage.}
\label{afm2}
\end{figure}

Small clusters (electron traps) can be realized, e.g., as small crystalline inclusions in the amorphous film, see Fig.~\ref{afm2}. It demonstrates inclusions of a round shape (marked by arrows) in topography AFM scan of the film area between the Ohmic contacts. These inclusions have sharp boundaries, they appear by 2 nm above the surface in profile.  The inclusions can be regarded as clusters of the crystalline phase in the amorphous film~\cite{butenko} and are of 0,5-1,5$\mu$m in size. 

\section{Conclusion}

In conclusion, we study  conductivity of strongly disordered amorphous antimony films  under high bias voltages.  We observe non-linear current-voltage characteristic, where the conductivity value at  zero bias is one of two distinct values, being determined by the sign of previously applied voltage. Relaxation curves demonstrate high stability of these  conductivity values on a large timescale. Investigations of the antimony film structure allows to determine the percolation character of electron transport in  strongly disordered films. We connect the memory effect in conductivity with modification of the percolation pattern due to recharging of some film regions at high bias voltages.

\section{Acknowledgement}
The authors are grateful to A.M. Ionov for fruitful discussions, S.V. Chekmazov for technical assistance, E.Yu. Postnova for electron microscopy, S.S. Khasanov for X-ray sample characterization, and V.M. Chernyak for preparation of amorphous films. We gratefully acknowledge financial support partially by the RFBR  (project No.~19-29-03021),  and RF State task.

\end{document}